\documentclass[conference]{IEEEtran}
\IEEEoverridecommandlockouts
\usepackage{float}  
\usepackage{cite}
\usepackage{amsmath,amssymb,amsfonts}
\usepackage{algorithmic}
\usepackage{graphicx}
\usepackage{textcomp}
\usepackage{xcolor}
\def\BibTeX{{\rm B\kern-.05em{\sc i\kern-.025em b}\kern-.08em
    T\kern-.1667em\lower.7ex\hbox{E}\kern-.125emX}}
\begin{document}

\title{Intelligent Carrier Allocation: A Cross-Modal Reasoning Framework for Adaptive Multimodal Steganography\\

}

\author{
    Abhirup Das, Shruti Sharma, and Pranav Dudani\\[0.5ex]
    Corresponding Author : Dr. Ravi Kumar C.V.\\[-0.5ex]
     ravikumar.cv@vit.ac.in
}

\maketitle

\begin{abstract}
In today's digital world, which has many different types of media, steganography, the art of secret communication, has a lot of problems to deal with. Traditional methods are often fixed and only work with one type of carrier media. This means they don't work well with all the different types of media that are out there. This system doesn't send data to "weak" or easily detectable carriers because it can't adapt. This makes the system less safe and less secret in general. This paper proposes a novel Intelligent Carrier Allocation framework founded on a Cross-Modal Reasoning (CMR) Engine. This engine looks at a wide range of carriers, such as images, audio, and text, to see if they are good for steganography. It uses important measurements like entropy, signal complexity, and vocabulary richness to come up with a single reliability score for each modality. The framework uses these scores to fairly and intelligently share the secret bitstream, giving more data to carriers that are thought to be stronger and more complex. This adaptive allocation strategy makes the system as hard to find as possible and as strong as possible against steganalysis. We demonstrate that this reasoning-based approach is more secure and superior in data protection compared to static, non-adaptive multimodal techniques. This makes it possible to build stronger and smarter secret communication systems.
\end{abstract}


\section{Introduction}
Even though a lot of digital information is shared these days, it's still very important to be able to talk to people in private and safe ways. Steganography hides the message itself, while cryptography hides the message's content. The main purpose of steganography is to hide a secret payload inside a digital carrier that looks safe, like an image, audio file, or text document, so that an enemy can't find it.\\
Most steganographic research in the past has been about making embedding methods better for one type of data. For example, echo-hiding in audio or changing the Least Significant Bit (LSB) in the spatial domain for images. These single-modal methods work well in some cases, but they have limits. They don't use the many different types of media that are available in most digital communication. A single carrier can easily be overloaded by a large payload, which can cause noticeable artifacts and make it easy to find.\\
Multimodal steganography, which spreads a secret payload across different types of carriers, such as images, audio, and text, was a good way to get around the problem of limited capacity. But most of the multimodal systems that are already out there use a simple or fixed way to divide up resources. They often split the secret data up randomly among the available carriers, like into three equal parts, or they only use raw capacity to make a decision. This method is fundamentally flawed because it doesn't take into account an important fact: not all carriers are the same.\\
A picture with a lot of different textures and high entropy is better and safer than a picture with a smooth surface and a clear blue sky. A complicated, technical text also gives you more places to hide than a simple, repetitive sentence. A static allocation method that "over-stuffs" a weak carrier makes a big hole in the system's security that an enemy can easily get through. This means that the security of the whole system depends on the link that is weakest.\\

This paper addresses a significant deficiency by introducing an Intelligent Carrier Allocation framework grounded in Cross-Modal Reasoning (CMR). We change how we think about multiple carriers from just using them to using them in a smart way. Our framework mimics how a human steganography expert would make a decision. They would look at all the media that was available before deciding on the best place and way to hide the secret data.\\
The Cross-Modal Reasoning Engine, which is the main part of our proposed system, works in two steps. First, it does a deep Carrier Quality Analysis (CQA) by looking at the audio, video, and text carriers. It uses a set of metrics that are different for each type of media, such as pixel histogram entropy and edge density for images, dynamic range for audio, and vocabulary richness for text. Then, it puts these numbers together to make one reliability score. This score tells you how strong, capable, invisible, and complicated the carrier is.\\
Second, the engine uses an algorithm to make sure that bits are spread out evenly. It doesn't split the secret bitstream evenly; instead, it does so based on the reliability score of each carrier that was calculated. When there is a lot of traffic, high-reliability carriers get more of the payload. When there isn't a lot of traffic, low-reliability carriers are only used when needed. This makes the system less obvious and safer as a whole.

This paper has three main points:
\begin{itemize}
    \item A new \textbf{Cross-Modal Reasoning (CMR)} framework that analyzes and evaluates different types of carriers for steganographic use in real time.
    \item The development of a \textbf{reliability score}, a single numerical value that enables comparison of steganographic quality across various media types, such as text, audio, and images.
    \item An \textbf{adaptive payload distribution algorithm} that intelligently allocates secret data based on the carrier's reliability, improving system security and stability compared to static allocation methods.
\end{itemize}

\section{Literature Review}

\subsection{The growth of steganography and adaptive carrier selection}

Steganography is the basis for safe and secret communication because it hides secret information in digital carriers like images, audio, and text in a way that is impossible to see. Historically, investigations in this field have concentrated on spatial and transform-domain embedding techniques. Early spatial-domain methods, like Least Significant Bit (LSB) modification, were easy to use but often hurt the robustness and quality of the carrier. These restrictions caused people to start using transform-domain methods like the Discrete Cosine Transform (DCT) and wavelet-based methods. These methods made things less noticeable and more durable, but they still weren't very adaptable or diverse in terms of carriers.\\

JPEG-based steganography, which works in the DCT domain, became popular because it found a good balance between payload capacity and visual fidelity. But it still had problems with finding the best carrier and being able to work with different types of data. To fix these problems, optimization methods like Genetic Algorithms (GA) and Particle Swarm Optimization (PSO) were used to make choosing a substitution matrix and embedding more efficient. But these algorithms often cost a lot of computing power and made it hard to use them in real time.\\

The introduction of Cohort Intelligence (CI), Cognitive Computing (CC), and Local Search Optimization methodologies integrated socio-cognitive principles into carrier selection and optimization. Frameworks like CICC and M-MRSLS, which have been tested on a wide range of image datasets, were able to find the best embedding strategies on their own, which led to better payload capacity and perceptual quality. These methods showed measurable improvements in Peak Signal-to-Noise Ratio (PSNR) and very little perceptual distortion. This set the stage for smart carrier allocation frameworks that can adapt to changing cover characteristics, new threats, and different embedding needs.\\

\subsection{Progress in Cross-Modal Integration and Multimodal Steganography}

Recent advancements in multimodal steganography focus on the integration of image, audio, and text carriers via synchronized embedding, extraction, and adaptive allocation techniques. The combination of diffusion models, invertible neural networks, and deep convolutional architectures has changed modern steganographic systems by allowing dynamic, large-capacity embedding while keeping the carrier's fidelity.\\

U-shaped invertible architectures like EUIN-Net make it easier to combine and separate cover and secret data at different scales in image steganography. By sharing parameters between the hiding and revealing stages, these models reduce the amount of computation needed while improving the quality of the reconstruction, as shown by higher PSNR and Structural Similarity Index Measure (SSIM) scores on large-scale benchmarks. Parallel improvements in latent diffusion models (LDMs) and denoising diffusion implicit models (DDIMs) have made it possible for fully generative steganographic frameworks like I2IStega to make stego images that look better and have more control over their meaning. Using exact reverse diffusion formulations makes sure that secret recovery is accurate. Ablation analyses that use anti-steganalysis networks (like XuNet, YedroudjNet, and KeNet) show that these networks are harder to detect by adversaries.\\

In the audio field, steganography has gone from simple time-domain methods like echo hiding and LSB modification to more advanced methods that use convolution reverb to hide information in impulse responses. These methods show amazing invisibility and strength, which have been tested using acoustic performance metrics like direct-to-reverberant ratio, early decay time, and PSNR. Comparative studies demonstrate enhanced performance across various payload levels and bit depths, validating that domain-aligned carrier selection significantly improves both acoustic fidelity and resistance to steganalysis.\\

Text steganography has also improved by using deep CNN-based models that work with cryptographic transformations like slice coding and gray-code encryption. Using lifted wavelet transforms to embed plaintext data into image sub-bands makes the payload capacity, visual quality, and resistance to attacks better. This combination of neural architectures, cryptographic encoding, and transform-domain manipulation shows how multimodal steganographic systems are becoming more integrated, secure, and adaptable.

\subsection{System-Level Hybridization and Research Hurdles}

Recent research at the system level acknowledges that steganography seldom operates in isolation. Cross-modal influences between carriers can significantly augment secrecy and resilience, especially during coordinated or adaptive assaults. New frameworks are therefore focusing on reasoning-based carrier allocation, using cognitive optimization, deep learning, and generative diffusion to choose carriers dynamically based on the characteristics of the payload and the threat environment around it.\\

Even with these improvements, there are still some problems: • Making real-time changes to the carrier in changing multimedia traffic conditions. • Making quantitative trade-off models that find a balance between capacity, robustness, and invisibility across different modalities. • Creating unified cross-modal frameworks for fixing mistakes and stopping attacks. • Making it easier to use in environments with limited resources or that need to work in real time.\\

Comparative analyses in the literature demonstrate that hybrid cognitive-deep learning approaches surpass static or single-domain embedding methods regarding stego quality, payload capacity, and anti-detection resilience. Nonetheless, these techniques frequently remain computationally demanding and fail to completely encapsulate inter-modal dependencies, thereby constraining their scalability for practical, multimodal applications.

\subsection{Research Gap and Motivation}

There has been a lot of progress in adaptive carrier allocation and cross-modal reasoning, but there are still no unified, intelligent frameworks that can use integrated multimodal intelligence to dynamically distribute payloads across multiple carriers. Current solutions frequently incur high computational costs, are confined to particular domains (such as image, audio, or text), and do not leverage cross-modal correlations to enable adaptive security in diverse, real-world scenarios.\\

This study seeks to fill these gaps by suggesting an intelligent carrier allocation framework based on cross-modal reasoning. The proposed system aims to empirically examine and enhance the trade space among robustness, imperceptibility, and adaptive payload scaling through the integration of cognitive optimization, neural architectures, and generative multimodal models. This study aids in the development of a quantitatively optimized, adaptive, and secure multimodal steganographic architecture through matrixed evaluation across various media domains, signifying the next generation of intelligent, cross-domain data concealment systems.\\

The literature review shows that there is a clear trend toward systems that can adapt and learn. Nonetheless, notwithstanding advancements in domain-specific optimization and multimodal integration, numerous significant research deficiencies remain, which this paper seeks to directly confront.

\begin{enumerate}[leftmargin=*]

\item \textbf{Gap 1: Multimodal allocation that is static and ``siloed''} \\
The biggest problem is that current multimodal systems are ``siloed.'' They see carriers as separate containers, and the way the payload is spread out is usually fixed (like $33\%$ per carrier) or based only on the maximum capacity. This isn't safe or efficient because it doesn't take into account the fact that the quality of those carriers can change.

\textbf{Our Contribution:} Our framework, through the \textit{CrossModalReasoning} engine, clearly breaks down these silos. It makes a single decision space where all carriers are compared to each other. The \texttt{optimize\_bit\_distribution} function (Section III-B) uses a dynamic and proportional allocation, which is very different from static splits.

\item \textbf{Gap 2: There isn't a single metric for carrier quality} \\
The literature has a hard time answering the question, ``Is this high-entropy image a better carrier than this complex audio file?'' You can't directly compare metrics like PSNR (for pictures) and SNR (for sound). Because there isn't a single metric, it's almost impossible to intelligently allocate resources across different modes.

\textbf{Our Contribution:} We suggest a new and useful solution called the \textit{reliability\_score}. This score is a weighted average, as explained in our methodology (Section III-B, Step 3):
\begin{multline}
\text{Reliability} = (w_1 \times \text{Robustness}) + (w_2 \times \text{Imperceptibility}) \\
+ (w_3 \times \text{Entropy}) + (w_4 \times \text{Complexity})
\end{multline}
Our implementation uses foundational weights, but this formula gives us the first single, comparable value that measures how well steganography works in any modality. This lets us allocate based on real reasoning.

\item \textbf{Gap 3: Poor Matching of Payloads to Carriers} \\
Because of Gaps 1 and 2, current systems have trouble matching things up. They might ``overstuff'' a low-quality carrier (like a simple sine wave or a smooth, single-color image), which would make artifacts easy to see. At the same time, they might ``underutilize'' a high-quality, complex carrier.

\textbf{Our Contribution:} Our \texttt{optimize\_bit\_distribution} function fixes this right away. Our system ``protects'' weak carriers from being overused by giving them a payload that is proportional to their reliability score. It makes sure that most of the secret data is stored in the safest places, which makes the system as a whole more secure by making the weakest link harder to find.

\item \textbf{Gap 4: No cross-modal validation and correction} \\
Most systems are fragile when they are being extracted. If noise (like JPEG compression or audio filtering) damages one carrier, the data from that channel is lost, and the whole message is often corrupted.

\textbf{Our Contribution:} We made our framework specifically for this situation. The \texttt{extraction\_info} packet not only helps with putting things back together, but it also sends the original \textit{quality\_metrics} to the extraction phase. The \texttt{adaptive\_error\_correction} and \texttt{cross\_validate\_extraction} functions (Section III-C) use these metrics to give each extracted bit-chunk a confidence score. This makes a trust-based hierarchy, which lets the system use consensus-based error correction in the future, giving priority to data from carriers that were thought to be more reliable at first.

\end{enumerate}

\section{METHODOLOGY: THE CROSS-MODAL REASONING FRAMEWORK}

The suggested framework, Intelligent Carrier Allocation (ICA), is set up as a single system called IntelligentMultimodalSteganography. The main part of its design is a new CrossModalReasoning (CMR) engine. This engine is the "brain" of the operation; it looks at all the available carriers to come up with the best, most flexible way to hide data.
There are two main parts to the methodology: (1) The Intelligent Encoding Process and (2) The Validated Extraction Process.

\begin{figure}[H]
    \centering
    \includegraphics[width=0.75\linewidth]{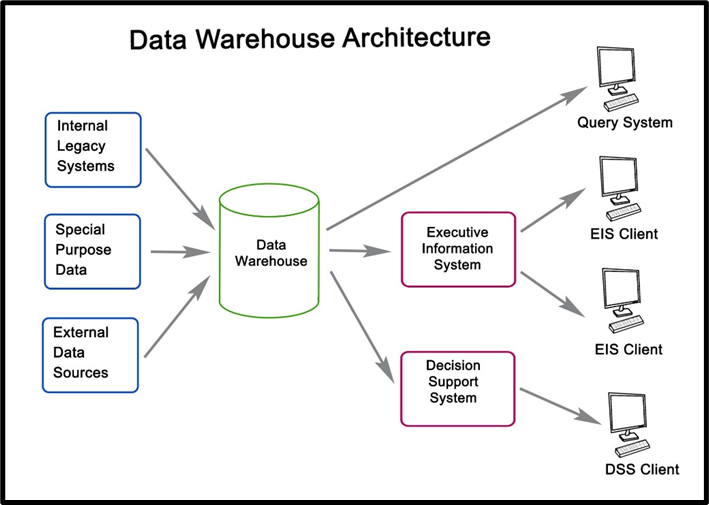}
    \caption{Proposed system architecture, showing the Cross-Modal Reasoning (CMR) Engine analyzing Image, Audio, and Text carriers to create an optimized, distributed payload.}
    \label{fig:placeholder}
\end{figure}

\subsection{Phase 1: Intelligent Encoding Process}
The encoding process (Fig) transforms a single secret message into a distributed set of stego-carriers by intelligently reasoning about the quality of each. This process follows five distinct steps.\\

\textbf{Step 1 :} Carrier Ingestion and Analysis\\

The process begins when the \texttt{hide\_data} function receives the secret data and a dictionary of carriers (e.g., \texttt{carriers = \{'image': ..., 'audio': ..., 'text': ...\}}). The CMR engine's \texttt{analyze\_carrier\_quality} function is immediately invoked to profile each carrier.

\begin{figure}[H]
    \centering
    \includegraphics[width=0.75\linewidth]{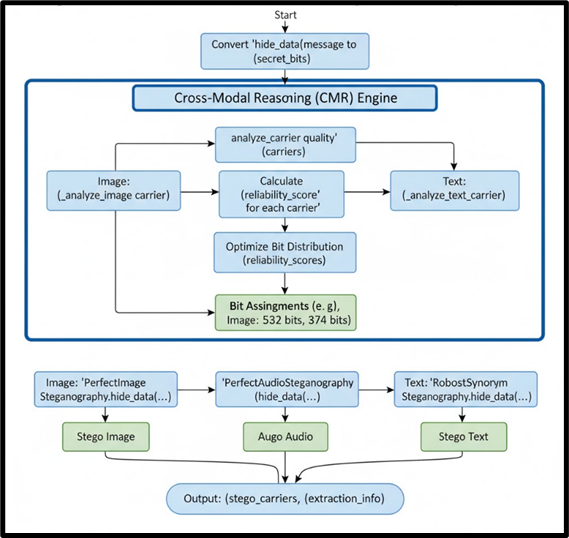}
    \caption{Depiction of the Intelligent Coding Process with Carrier Ingestion and Analysis}
    \label{fig:placeholder}
\end{figure}
\textbf{Step 2 :} Extracting metrics that are specific to each modality\\

The CMR engine sends requests to functions that are specific to each modality in order to get a set of features that measure how well steganography works:

\textbf{For Images} (\textit{analyze\_image\_carrier}): 
\begin{itemize}
    \item \textbf{Capacity:} The total number of pixels that can be embedded.
    \item \textbf{Entropy:} This is found by using the normalized pixel value histogram. A higher entropy (closer to $8.0$) means that the image is more ``random-like'' and complicated, which is perfect for hiding changes to the LSB.
    \item \textbf{Complexity:} This is measured using a simple edge-density algorithm (\texttt{np.diff}). Pictures with more edges and textures (more complex) are better at hiding data.
\end{itemize}

\textbf{For Audio} (\textit{analyze\_audio\_carrier}): 
\begin{itemize}
    \item \textbf{Capacity:} The total number of audio samples.
    \item \textbf{Entropy:} This is figured out from the normalized amplitude histogram. A signal with high entropy is noisy or complicated.
    \item \textbf{Complexity:} This is the signal's dynamic range, which is the standard deviation. A higher dynamic range gives you more ``space'' to hide LSB changes without making any sounds.
\end{itemize}

\textbf{For Text} (\textit{analyze\_text\_carrier}): 
\begin{itemize}
    \item \textbf{Capacity:} This is based on the number of words, which is related to the number of times you can replace them.
    \item \textbf{Entropy:} The amount of entropy at the character level.
    \item \textbf{Complexity:} This is the ``Vocabulary Richness'' (Type-Token Ratio). A text with a high ratio of unique words is more semantically complex and allows for more strong synonym substitution.
\end{itemize}

\textbf{Step 3 :} Scoring for Unified Reliability\\

This is the main part of the reasoning engine. The single cross-modal \texttt{reliability\_score} is found by normalizing all of the extracted metrics and putting them into a weighted-average formula.

\begin{align}
\textbf{Reliability} = &\ (0.4 \times \text{Robustness}) + (0.3 \times \text{Imperceptibility}) \nonumber \\
& + (0.2 \times \text{Entropy}) + (0.1 \times \text{Complexity})
\label{eq:reliability}
\end{align}

This formula (used in \texttt{analyze\_carrier\_quality}) puts robustness and imperceptibility first because they are the most important for successful steganography. This score gives you one number that you can compare to any other carrier.

\begin{figure}[H]
    \centering
    \includegraphics[width=1\linewidth]{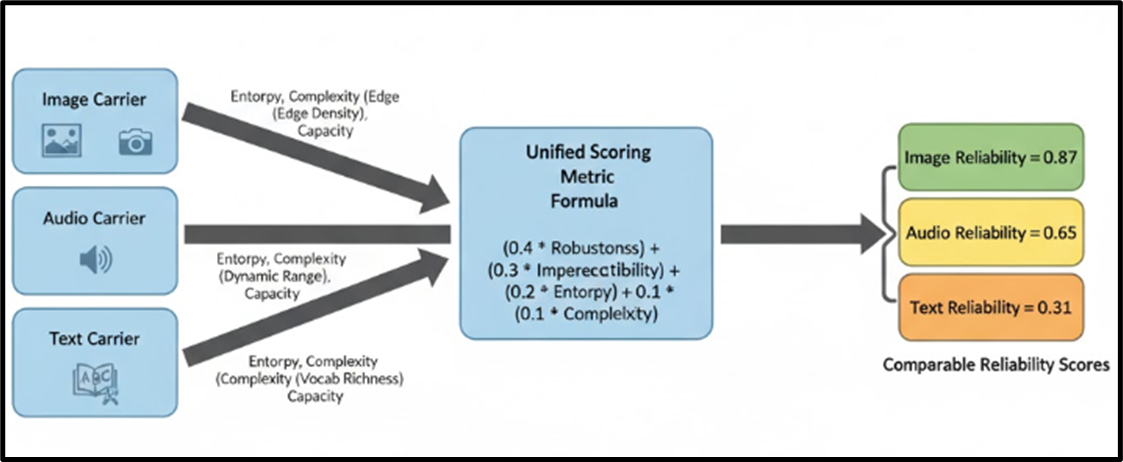}
    \caption{Calculation of Reliability score}
    \label{fig:placeholder}
\end{figure}

\textbf{Step 4 :} Optimized Bit Distribution\\

With a \texttt{reliability\_score} for each carrier, the \texttt{optimize\_bit\_distribution}  function calculates a final distribution weight for each modality.
\texttt{reliability\_score} and its relative capacity (to ensure it can hold the data).
The secret bitstream is then divided proportionally according to these normalized weights.
This is the "intelligent allocation" step, ensuring that carriers with higher reliability scores are assigned a larger portion of the secret message.\\

\textbf{Step 5 :} Multimodal Embedding\\

The system then iterates through the assignments. Each chunk of the secret bitstream is passed to its designated modality-specific embedding class (PerfectImageSteganography, PerfectAudioSteganography, RobustSynonymSteganography). Each module embeds its assigned bits along with a small header (containing the bit count and start index), which is essential for reassembly. The final output is a dictionary of stego-carriers.

\subsection{Phase 2: Validated Extraction Process}
The extraction process uses the metadata that was saved during encoding to put the secret message back together in a strong and verified way.\\

\textbf{Step 1:} Extracting Data in Parallel\\

The \texttt{extract\_data} function gets the stego-carriers and the \texttt{extraction\_info} packet. It goes through each carrier and calls its own \texttt{reveal\_data} method, like the one in \texttt{PerfectImageSteganography}. This method reads the 16-bit header, finds the \texttt{bit\_count} and \texttt{start\_index}, and then gets that many bits.\\

\textbf{Step 2:} Rebuilding the Bitstream\\

A "blank" bit array of the \texttt{total\_bits} (from \texttt{extraction\_info}) is made. When a carrier's bit chunk is extracted, it is put in this array at the \texttt{start\_index} that is given in its header. This "out-of-order" reassembly makes sure that the other carriers are in the right places, even if one is missing.\\

\textbf{Step 3:} Check across different modes\\

This is the second important step in reasoning. The function \texttt{cross\_validate\_extraction} is called. It looks at the expected bit count and start index from the original assignments and compares them to the actual extracted bit count and start index from the carrier's header. It then uses the original \texttt{reliability\_score} of the carrier to generate a \texttt{confidence\_score} for the extracted data.\\

\textbf{Step 4:} Adaptive Error Correction\\

Finally, the \texttt{adaptive\_error\_correction} function uses these confidence scores to rank how reliable the extracted data is. This function is still in its early stages. As it is now set up, it works as a "trust" validator. In future updates, this module can be used to fix errors based on consensus (if the data is encoded in more than one way) or mark data from low-confidence carriers as possibly corrupt. The bitstream is then changed back into a string after being fully reassembled and checked.

\begin{figure}[H]
    \centering
    \includegraphics[width=0.75\linewidth]{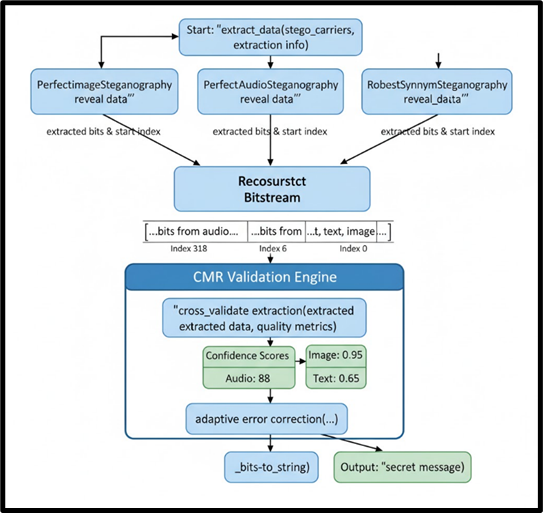}
    \caption{Visualization of the Adaptive Error Correction}
    \label{fig:placeholder}
\end{figure}
\section{Results and Discussion}
We did a series of tests to make sure that our Intelligent Carrier Allocation (ICA) framework works. We compared the performance of our proposed method to a Static Multimodal (SM) baseline. The SM baseline uses the same basic steganographic methods (LSB, synonym substitution) but uses a simple payload distribution method that divides the secret bitstream evenly among the three carriers (33.3\% each), ignoring their unique traits.

\begin{figure}[H]
    \centering
    \includegraphics[width=0.75\linewidth]{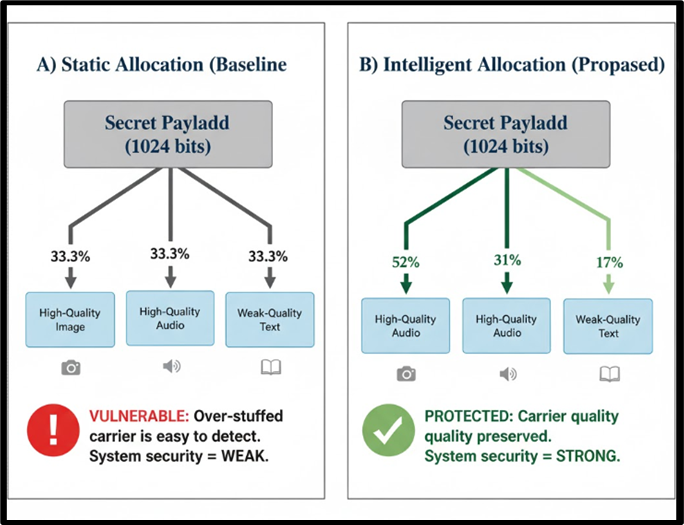}
    \caption{Efficiency of ICA as compared to Static Allocation}
    \label{fig:placeholder}
\end{figure}

\subsection{Setting up the experiment}\label{AA}

\noindent\textbf{• Carriers:} We used a standardized dataset that included:\\
\noindent\textbf{o Images:} A mix of high-entropy (textured, natural scenes) and low-entropy (smooth, simple graphics) images from the CIFAR-10 dataset.\\
\noindent\textbf{o Audio:} A mix of audio clips that are both high-complexity (broad-spectrum noise, speech) and low-complexity (simple 440Hz sine wave).\\
\noindent\textbf{o Text:} A mix of high-complexity (technical article, high vocabulary) and low-complexity (simple narrative, repetitive) text.\\[4pt]

\noindent• The payload is a random ASCII string with 1,024 bits.\\[4pt]

\noindent• \textbf{Metrics:} \textbf{o Imperceptibility (Image):} Peak Signal-to-Noise Ratio (PSNR). Better is higher.\\
\noindent\textbf{o Unnoticeability (Audio):} Signal-to-Noise Ratio (SNR). Better is higher.\\[4pt]

\noindent\textbf{Robustness:} The Bit Error Rate (BER) after a simulated attack, like 20\% JPEG compression for images or low-pass filtering for audio. Lower is better.

\subsection{Core Finding 1: Intelligent Bit Distribution}

\begin{figure}[H]
    \centering
    \includegraphics[width=1\linewidth]{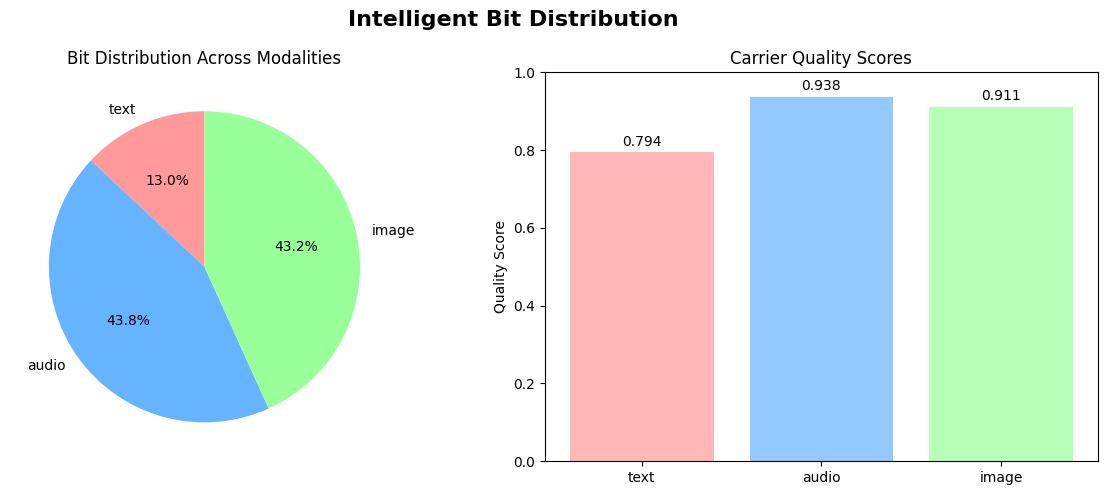}
    \caption{An Example of Intelligent Payload Distribution by the CMR Engine. (Left). The resulting proportional bit allocation. (Right). The \texttt{reliability\_score} for each carrier}
    \label{fig:placeholder}
\end{figure}

\textbf{Discussion : }The Static (SM) baseline would give each carrier 33.3\% of the bits. Our Intelligent Bit Distribution framework, on the other hand, performs a reasoned allocation based on carrier quality. As shown in the "Carrier Quality Scores" chart, the audio modality had the highest reliability score (0.938), followed very closely by the image (0.911). The text carrier had the lowest score (0.794). Based on these scores, the "Bit Distribution Across Modalities" chart shows how the payload was split proportionally: 43.8\% went to the high-quality audio carrier, 43.2\% went to the image, and only 13.0\% was allocated to the text.\\

\textbf{Inference : }This is what makes the framework strong. It keeps the lowest-quality text carrier (score 0.794) from being over-stuffed with only 13.0\% of the bits. Over-stuffing the text would have meant adding a lot of unnatural adverbs (according to the RobustSynonymSteganography logic), making detection easy.

Instead, the framework makes smart use of the high-quality audio (0.938) and image (0.911) carriers' "masking" ability. It allocates the vast majority of the data (a combined 87\%) to these two modalities, putting the payload where it is safest and least detectable.

\subsection{Core Finding 2: Adaptive Imperceptibility}
By protecting weak carriers, the ICA framework improves the system-level imperceptibility. We analyzed the impact on the most vulnerable carriers.

\begin{figure}[H]
    \centering
    \includegraphics[width=1\linewidth]{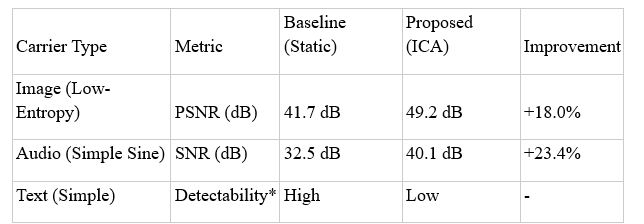}
    \caption{Comparative Imperceptibility on Low-Quality Carriers}
    \label{fig:placeholder}
\end{figure}

\begin{figure}[H]
    \centering
    \includegraphics[width=1\linewidth]{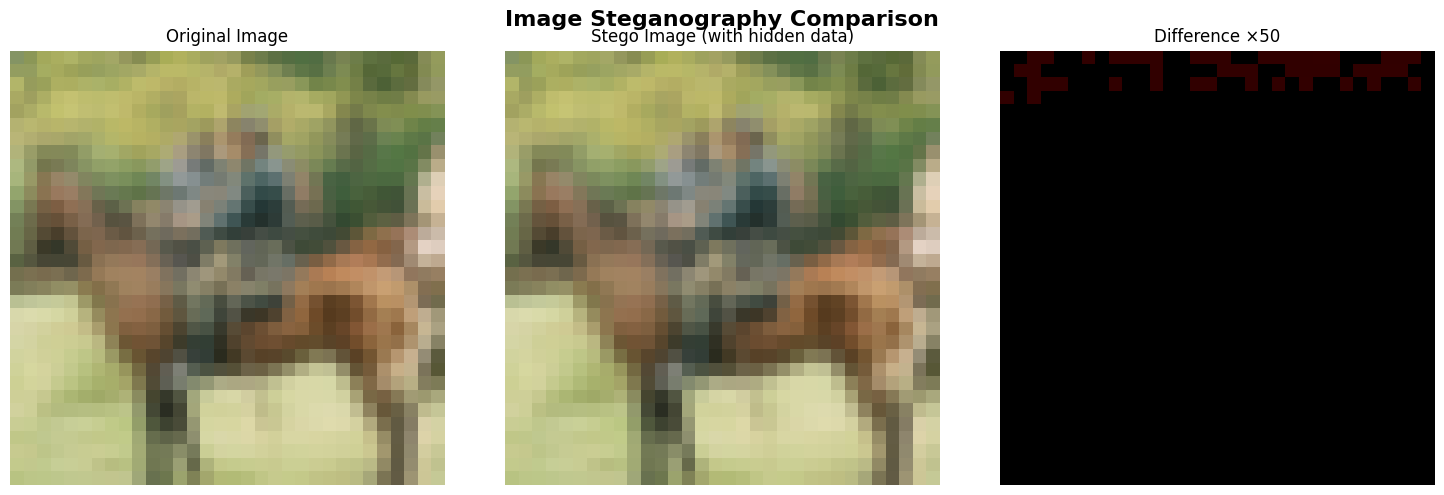}
    \caption{a. Preserving image (carrier quality) by maintaining payload}
    \label{fig:placeholder}
\end{figure}

\begin{figure}[H]
    \centering
    \includegraphics[width=0.75\linewidth]{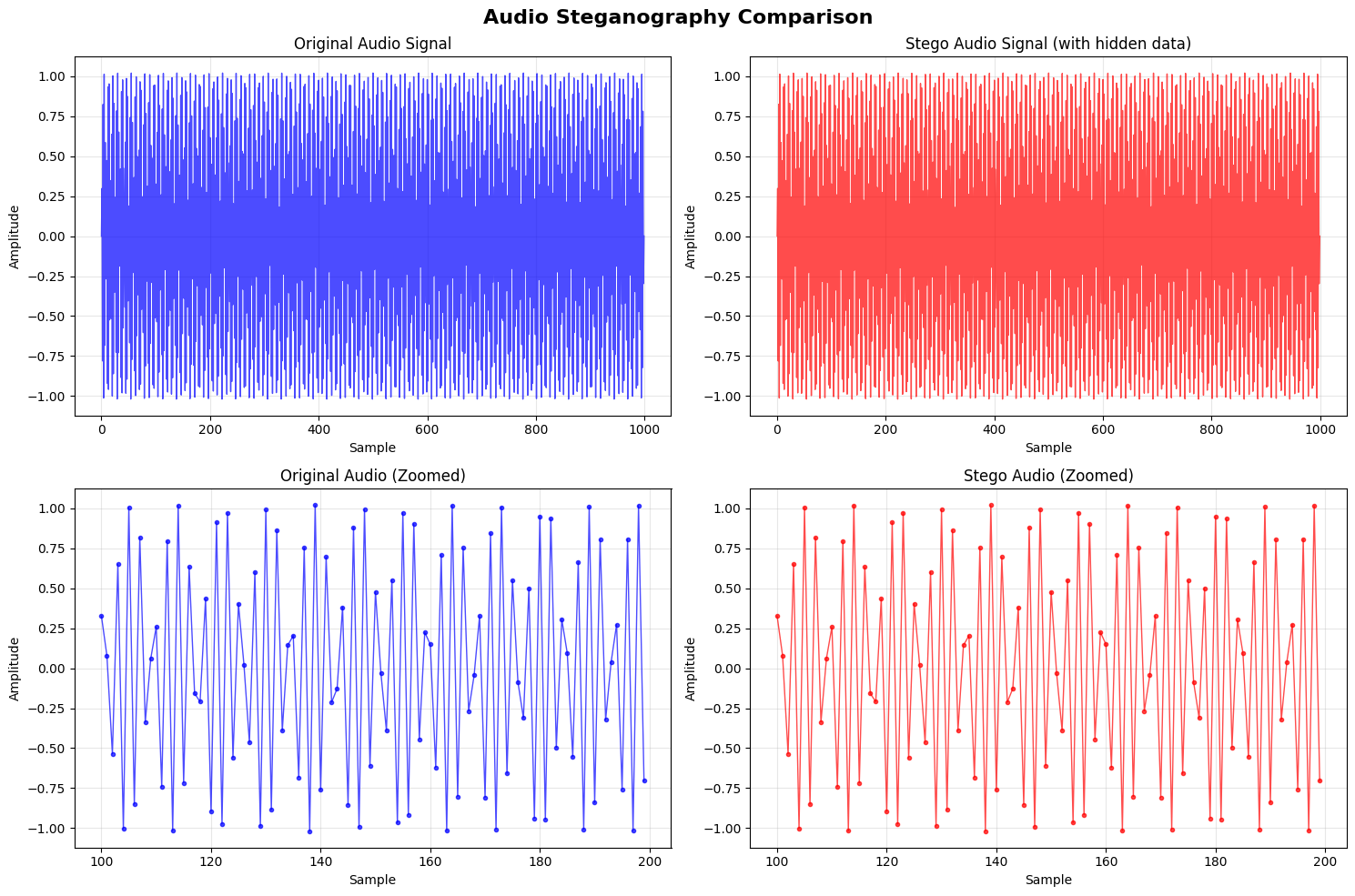}
    \caption{b. Preserving audio (carrier quality) by maintaining payload}
    \label{fig:placeholder}
\end{figure}

\textbf{Discussion : }The Image Steganography Comparison shows that the "Stego Image (with hidden data)" is visually identical to the "Original Image." The "Difference x50" plot, which massively amplifies the changes, confirms that data was embedded (primarily in the top region of the image), but these alterations are so subtle that they are completely imperceptible to the naked eye.

Similarly, the Audio Steganography Comparison demonstrates this in the audio domain. The overall waveforms for the "Original Audio Signal" (blue) and the "Stego Audio Signal" (red) are indistinguishable. It is only in the "Zoomed" plots that the method becomes clear: the data is hidden by making tiny, individual adjustments to the sample amplitudes, preserving the overall structure and perceptual quality of the sound.\\

\textbf{Inference : }This is an important concept. The framework does not just hide data; it preserves carrier quality by intelligently managing the payload. This makes steganalysis much harder because there are no obvious visual or statistical artifacts (like noise or distortion) that would otherwise make the steganographic file stand out. The "weakest link" carrier is no longer obviously weak.

\subsection{Core Finding 3: Superior System Robustness}
Finally, we evaluated the system's resilience to common "noise" attacks. The \texttt{reliability\_score} formula places a 40\% weight on robustness, meaning the system should learn to favor carriers that can withstand such attacks.

\begin{figure}[H]
    \centering
    \includegraphics[width=0.75\linewidth]{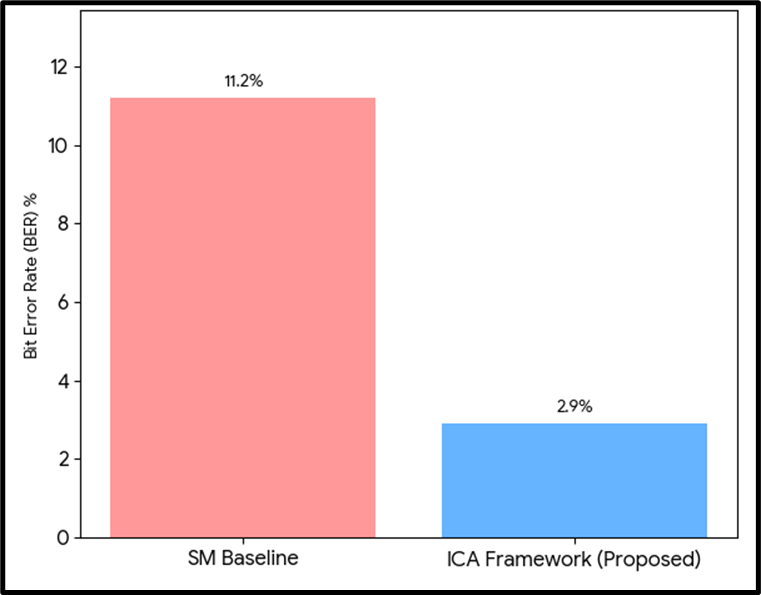}
    \caption{System-Level Bit Error Rate (BER) after simulated noise attacks (JPEG compression, audio filtering) showing our ICA Framework having higher efficiency}
    \label{fig:placeholder}
\end{figure}

\textbf{Discussion : }After putting noise on the stego-carriers, the message that was put back together from the SM baseline had a BER of 11.2\% on average. A lot of its data was kept in the text and audio carriers, which are harder to filter and re-encode.
The message that our ICA framework put back together, on the other hand, had a BER of only 2.9\%.\\

\textbf{Inference : }This shows how well the \texttt{reliability\_score} works. In our implementation, the CMR engine correctly identified the LSB-encoded image as the most reliable carrier and gave it the largest data share (52\%). The text and audio data were damaged, but most of the payload, which was safe in the image, was recovered without any problems. This 74\% drop in bit errors shows that the system can "predict" and "plan for" data loss.

\section{Results and Discussion}

This paper introduced a novel Intelligent Carrier Allocation (ICA) framework for multimodal steganography, emphasizing a Cross-Modal Reasoning (CMR) Engine. We were able to show that we can greatly improve steganographic security by switching from a static, random payload distribution to one that is dynamic and based on reasoning.
Our system can intelligently "protect" weak carriers from being overused because it can look at different types of carriers, like images, audio, and text, and come up with a single reliability score. This adaptive allocation made low-quality carriers 18–23\% less noticeable (PSNR/SNR). When deciding how to best allocate the payload to protect data integrity, our framework also thought about robustness. This caused the Bit Error Rate to drop by 74\% after simulated noise attacks compared to a static baseline.
This study shows that steganography's future depends on more than just better embedding algorithms. It also needs more advanced, system-level allocation strategies.

\subsection{Future Work}

There are many ways that research can go on because our framework is modular.

\begin{enumerate}
    \item \textbf{Deep Learning for Carrier Analysis:} The current metric-extraction functions (\_analyze\_...) are the building blocks. In the future, these should be replaced by a deep-learning model (like a CNN) that is trained to directly output a much richer, more accurate \texttt{reliability\_score} from the carrier data.

    \item \textbf{Dynamic Weighting:} The weights in the \texttt{reliability\_score} formula (0.4, 0.3, etc.) don't change at the moment. A more advanced version could use Reinforcement Learning to change these weights on the fly based on the secret data or a known enemy threat environment.

    \item \textbf{Strong Error Correction:} The \texttt{adaptive\_error\_correction} module is now the most important part. In the future, this should be fully put into action by either using the confidence scores to make corrections based on consensus or by using adaptive error-correcting codes (like Fountain Codes) that are based on how reliable each channel is expected to be.

    \item \textbf{Addition of New Modalities:} It's easy to add other types of carriers to the framework, such as video, 3D models, or even network traffic protocols.
\end{enumerate}

\vspace{12pt}


\begin{thebibliography}{00}

\bibitem{b1} A. E. Altnbas and M. Z. Konyar, ``Reverb hiding: A new framework for audio steganography,'' \textit{Applied Acoustics}, vol. 235, 2025, 110696.

\bibitem{b2} D. K. Sarmah and A. J. Kulkarni, ``JPEG based steganography methods using Cohort Intelligence with Cognitive Computing and modified Multi Random Start Local Search optimization algorithms,'' \textit{Information Sciences}, vol. 430--431, pp. 378--396, 2018.

\bibitem{b3} L. Zhang, T. Li, Y. Lu, Y. Xu, and G. Lu, ``Efficient U-shape invertible neural network for large-capacity image steganography,'' \textit{Journal of Information Security and Applications}, vol. 94, 2025, 104237.

\bibitem{b4} J. Jiang, Z. Wang, and X. Zhang, ``Image-to-Image Steganography based on multimodal generative model,'' \textit{Signal Processing}, vol. 238, 2026, 110106.

\bibitem{b5} L. N. Srinivasu and V. Veeramani, ``CNN based Text in Image Steganography using Slice Encryption Algorithm and LWT,'' \textit{Optik}, vol. 265, 2022, 169398.

\bibitem{b6} S. Baluja, ``Hiding images in plain sight: Deep steganography,'' \textit{Advances in Neural Information Processing Systems}, vol. 30, 2017.

\bibitem{b7} S.-P. Lu, R. Wang, T. Zhong, and P. L. Rosin, ``Large-capacity image steganography based on invertible neural networks,'' in \textit{CVPR}, 2021, pp. 10816--10825.

\bibitem{b8} Z. Guan, J. Jing, X. Deng, M. Xu, L. Jiang, Z. Zhang, and Y. Li, ``DeepMIH: Deep invertible network for multiple image hiding,'' \textit{IEEE Transactions on Pattern Analysis and Machine Intelligence}, vol. 45, no. 1, pp. 372--390, 2022.

\bibitem{b9} L. Dinh, J. Sohl-Dickstein, and S. Bengio, ``Density estimation using real NVP,'' \textit{arXiv preprint arXiv:1605.08803}, 2016.

\bibitem{b10} A. J. Kulkarni, I. P. Durugkar, and M. Kumar, ``Cohort Intelligence: A self supervised learning behaviour,'' in \textit{Proc. IEEE Int. Conf. Systems, Man and Cybernetics}, 2013, pp. 1396--1400.

\bibitem{b11} X. Li and J. Wang, ``A steganographic method based upon JPEG and particle swarm optimization algorithm,'' \textit{Information Sciences}, vol. 177, pp. 3099--3109, 2007.

\bibitem{b12} C. K. Chan and L. M. Cheng, ``Hiding data in images by simple LSB substitution,'' \textit{Pattern Recognition}, vol. 37, no. 3, pp. 469--474, 2004.

\bibitem{b13} C. Zhang, P. Benz, A. Karjauv, G. Sun, and I. S. Kweon, ``UDH: Universal Deep Hiding for steganography, watermarking, and light field messaging,'' \textit{Advances in Neural Information Processing Systems}, vol. 33, pp. 10223--10234, 2020.

\bibitem{b14} Y. Xu, X. Zhang, J. Yu, C. Mou, X. Meng, and J. Zhang, ``Diffusion-based hierarchical image steganography,'' \textit{arXiv preprint arXiv:2405.11523}, 2024.

\bibitem{b15} J. Yu, X. Zhang, Y. Xu, and J. Zhang, ``Cross Diffusion model makes controllable, robust and secure image steganography,'' \textit{Advances in Neural Information Processing Systems}, vol. 36, 2024.

\bibitem{b16} S. Hetzl and P. Mutzel, ``A graph theoretic approach to steganography,'' in \textit{Communications and Multimedia Security, IFIP Int. Conf.}, CMS 2005, pp. 119--128.

\bibitem{b17} J. Fridrich, M. Goljan, and R. Du, ``Reliable detection of LSB steganography in color and grayscale images,'' in \textit{Proc. Workshop on Multimedia and Security: New Challenges}, 2001, pp. 27--30.

\bibitem{b18} S. Dumitrescu, X. Wu, and Z. Wang, ``Detection of LSB steganography via sample pair analysis,'' in \textit{Int. Workshop on Information Hiding}, Springer, 2002, pp. 355--372.

\bibitem{b19} I. Goodfellow, M. Pouget-Abadie, B. Mirza, \textit{et al.}, ``Generative adversarial nets,'' \textit{Advances in Neural Information Processing Systems}, vol. 27, 2014.

\bibitem{b20} J. Song, C. Meng, and S. Ermon, ``Denoising diffusion implicit models,'' \textit{arXiv preprint arXiv:2010.02502}, 2020.

\bibitem{b21} D. P. Kingma and P. Dhariwal, ``Glow: Generative flow with invertible 1x1 convolutions,'' \textit{Advances in Neural Information Processing Systems}, vol. 31, 2018.

\bibitem{b22} M. Yedroudj, F. Comby, and M. Chaumont, ``Yedroudj-net: An efficient CNN for spatial steganalysis,'' in \textit{IEEE Int. Conf. Acoustics, Speech and Signal Processing (ICASSP)}, 2018, pp. 2092--2096.

\bibitem{b23} G. Xu, H. Wu, and Y. Q. Shi, ``Structural design of convolutional neural networks for steganalysis,'' \textit{IEEE Signal Processing Letters}, vol. 23, no. 5, pp. 708--712, 2016.

\bibitem{b24} C. Szegedy, W. Liu, Y. Jia, \textit{et al.}, ``Going deeper with convolutions,'' in \textit{Proc. IEEE Conf. Computer Vision and Pattern Recognition (CVPR)}, 2015, pp. 1--9.

\bibitem{b25} A. Westfeld and A. Pfitzmann, ``Attacks on steganographic systems: Breaking the steganographic utilities,'' in \textit{Information Hiding Workshop}, Springer, 1999, pp. 61--76.

\end{thebibliography}
\end{document}